# ChatMOSP: A Chemistry-Grounded Mobile Agent for Working-State Catalyst Simulations

Sanyang Ye[1, 4], Rui Qi[2, 3*], Beien Zhu[2, 3*], and Yi Gao[2, 3*]

[1]*Key Laboratory of Interfacial Physics and Technology, Shanghai Institute of Applied Physics, Chinese Academy of Sciences, Shanghai 201800, China*

[2]*Photon Science Research Center for Carbon Dioxide, Shanghai Advanced Research Institute, Chinese Academy of Sciences, Shanghai 201210, China*

[3]*State Key Laboratory of Low Carbon Catalysis and Carbon Dioxide Utilization, Shanghai Advanced Research Institute, Chinese Academy of Sciences, Shanghai 201210, China*

[4]*University of Chinese Academy of Sciences, Beijing 100049, China*

*To whom correspondence should be addressed. Email: qir@sari.ac.cn; zhube@sari.ac.cn; gaoyi@sari.ac.cn

ABSTRACT: Catalytic nanoparticles restructure dynamically under reaction conditions, so their working morphology and activity are governed by temperature, pressure, and gas composition.

However, converting experimentally specified environments into physically meaningful morphology-performance simulations remains difficult because the translation of reaction conditions into model-specific energetic, kinetic, and execution parameters requires the specialized knowledge in computational catalysis. Here we report ChatMOSP, a chemistry-grounded mobile scientific agent that translates natural-language and voice-expressed catalytic requests into parameter-validated simulations using the Multi-scale Operando Simulation Package. ChatMOSP maps catalyst identity, temperature, pressure, gas composition, and target observables onto multiscale structure reconstruction and kinetic Monte Carlo tasks, retrieves database parameters or constructs missing inputs from an online literature-retrieval workflow, and executes validated MOSP workflows. Using CO oxidation on Pd nanoparticles as an example, we verify the ChatMOSP simulations capture the temperature-induced transition from faceted to rounded morphologies observed by in-situ TEM experiments either by built-in database or from web-retrieved literature information when the parameters are absent. Moreover, we demonstrate the capability of ChatMOSP to perform end-to-end study at mobile devices to simulate a pressure-coverage-morphology-activity feedback cycle for Pt CO oxidation to interpret the oscillatory CO conversion. These results establish ChatMOSP as a physically constrained mobile agent for accessible and interpretable catalyst working-state simulations.

## Introduction

Metal nanoparticles are central to heterogeneous catalysis, where their performance is closely linked to morphology because particle shape determines the exposed facets and available active sites. Under operando conditions, however, nanoparticles are not static structures but dynamic working states whose morphology and surface chemistry respond to temperature, pressure, and gas composition, thereby altering active-site distributions and catalytic rates.[1-3] In situ and operando experiments have shown that reactive gases such as CO, $O_2$, and NO can reshape metal nanoparticles and nanoalloys, underscoring the importance of probing catalyst structures under working conditions to understand their catalytic performance.[4-13] However, experimental characterization of such dynamic restructuring relies on advanced in situ techniques and remains costly, technically demanding, and difficult to extend systematically across diverse catalyst-reaction systems. Theoretical modeling is therefore essential for quantitatively linking reaction environments to nanoparticle morphology and catalytic behavior.

Physics-based morphology simulations provide an effective route to bridge this gap by resolving atomistic and mesoscale structural features that are difficult to access directly in experiments.[14-17] Hensen and co-workers related static or quasi-static Wulff-type nanoparticle morphology to catalytic activity using DFT energetics and microkinetics.[18, 19] Scaling-relation Monte Carlo methods developed by Grönbeck and co-workers have enabled important progress on predefined nanoparticle structures to resolve site-dependent kinetics under reaction conditions.[20, 21] Recently, our group developed the Multi-scale Operando Simulation Package (MOSP), which integrates multiscale structure reconstruction (MSR) and kinetic Monte Carlo (KMC) modules to directly link externally imposed reaction conditions with nanoparticle morphology, exposed facets, adsorbate populations, and catalytic performance.[22, 23] This framework enables reaction-dependent morphology-performance relationships to be analyzed beyond the direct reach of operando observation. However, its use still requires specialized knowledge of multiscale morphology and kinetic models. Experimentally familiar conditions, such as catalyst identity, temperature, pressure, gas composition, and target observables, must be translated into software-specific inputs, including energetic parameters, adsorbate corrections, kinetic settings, model assumptions, unit-consistent variables, and execution logic. This operational barrier limits the routine use of MOSP-type simulations by experimental and non-specialist users, motivating a more accessible interface for converting chemically expressed reaction conditions into executable, physically validated morphology-kinetics simulations.

Workflow automation and large-language-model-based scientific agents offer important foundations for bridging human intent and computational execution.[24-27] Automated workflow platforms (such as AiiDA, FireWorks, and QMflows) have improved execution control, provenance tracking, high-throughput calculation, and heterogeneous workflow coordination in computational chemistry and materials science,[28-32] while recent chemistry- and materials-oriented agents, including ChemCrow, Coscientist, and OpenClaw-based agent-skill frameworks, have demonstrated natural-language task formulation, modular planning, and domain-tool invocation.[33-39] Nevertheless, most existing systems either automate predefined computational procedures or coordinate general-purpose chemistry tools, which are incapable to address a specific catalytic science problem, for example, the catalyst structure and activity evolving within the reaction environment. For reaction-condition-dependent catalyst simulations, an effective agent must therefore do more than parse commands or launch calculations. It must translate chemically expressed reaction conditions into physically constrained model variables, retrieve or construct missing parameters, verify model consistency, and generate executable inputs while preserving the mechanistic connection among reaction atmosphere, adsorbate coverage, morphology evolution, and catalytic performance.

Here, we report ChatMOSP, a chemistry-grounded mobile scientific agent that connects experimentally expressed catalytic conditions with MOSP-based catalyst working-state simulations. Rather than presenting a general-purpose workflow assistant, we focus on a specific catalytic science problem: how reaction temperature and local gas pressure reshape catalytic nanoparticles and regulate their performance under CO oxidation conditions. ChatMOSP converts natural-language and voice-expressed requests into parameter-validated MSR and KMC simulations, using built-in database when parameters are available and web-based literature retrieval when parameters are missing. By using Pd nanoparticles under CO oxidation as example, ChatMOSP reproduces the experimentally observed temperature-induced faceting-to-rounding transition either from built-in database or from literature-derived information. To further testify its end-to-end full scale study capability, for Pt CO oxidation, ChatMOSP captures a CO-pressure-dependent Pt morphology transition from a faceted, low-activity state under CO-depleted conditions to a rounded, open-site-rich, high-activity state under CO-rich conditions, linking morphology evolution with KMC-derived activity differences to provide a physically interpretable feedback picture for experimentally observed oscillatory CO oxidation. This progression demonstrates that ChatMOSP is not merely a mobile interface to simulation software, but a chemically constrained route for linking experimental reaction environments with operando nanoparticle morphology and catalytic performance.

## Results and Discussion

### ChatMOSP Workflow for Model-Preserving MOSP Simulations

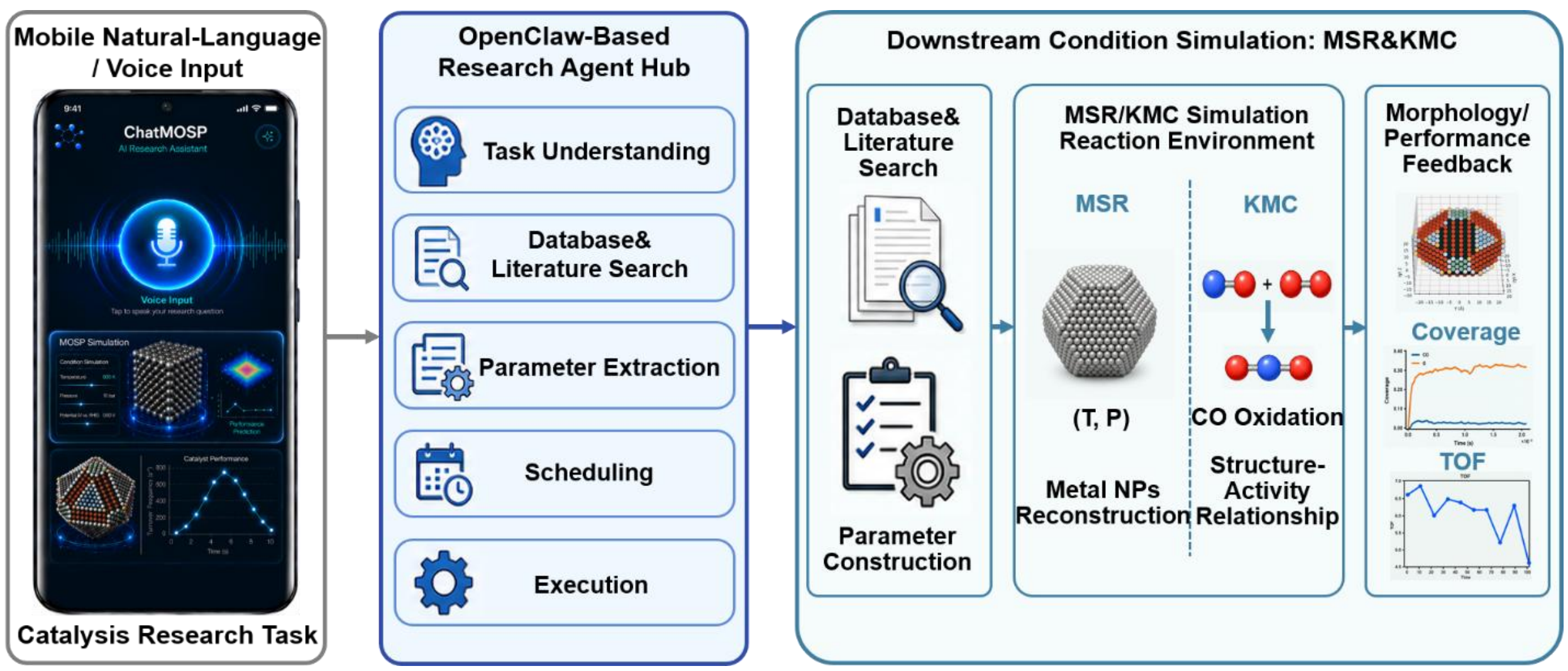

**Figure 1.** Closed-loop ChatMOSP workflow for mobile catalyst-state simulation.

The ChatMOSP framework was designed to translate experimentally expressed catalytic conditions into executable catalyst-state simulations rather than to act as a general chatbot or graphical wrapper (Figure 1). In the current implementation, users access ChatMOSP through the Feishu (https://www.feishu.cn/?from_site=lark) mobile interface and provide catalyst identity, reaction temperature, gas composition, pressure, and target observables by text or voice. An OpenClaw-based hierarchical agent, using GLM-5 (https://glm-5.org) for language understanding and task coordination, interprets the request and maps the chemically meaningful variables onto the MSR and KMC task schemas required by MOSP.

A central design principle of ChatMOSP is that the language model is restricted to task interpretation, workflow coordination, and parameter organization rather than direct scientific prediction. Tests with different language-model backbones showed that general language models alone do not provide physically validated nanoparticle morphologies, surface coverages, or kinetic outputs (Figure S1). These outputs become available only after the agent is equipped with MOSP-specific skills that construct validated inputs, execute MSR and KMC simulations, and return the resulting structures and kinetic descriptors. Accordingly, the workflow is organized through a hierarchical skill architecture, including a central controller and task-specific subskills for parameter extraction, database retrieval, literature-assisted parameter construction, MOSP input generation, MSR/KMC execution, and result visualization (Figure S2). This skill-mediated connection to MOSP is compatible with different language-model backbones, while all catalyst-state predictions are produced by the underlying MOSP physical models after the required surface energies, adsorption energies, adsorbate interactions, kinetic parameters, environmental variables, and execution settings have been validated.

To verify that the agent-mediated workflow preserves the original MOSP simulation capability, we tested ChatMOSP using the built-in Pt-CO/O parameter set established in our previous MOSP studies. ChatMOSP automatically retrieved the required energetic and interaction parameters, generated MOSP-compatible MSR inputs, and executed morphology reconstruction under the same reaction-condition settings (Figure S3 and S4). The resulting temperature-dependent Pt morphology evolution reproduced the trend previously reported by the original expert-operated MOSP workflow, confirming that ChatMOSP acts as a model-preserving interface rather than an independent structure generator.[40] Using this validation case, we further tested different language-model backbones, including DeepSeek-V3.2, MiniMax-M2.5, Kimi-K2.5, and GLM-5, within ChatMOSP. Consistent morphology outputs were obtained once the requests were mapped to the same MOSP inputs and parameter settings, demonstrating that the catalyst-state predictions are governed by the MOSP physical model rather than by the language-model backbone (Figure S5). Thus, ChatMOSP improves accessibility to MOSP simulations while preserving the underlying physical behavior.

**Database-Guided Reproduction of Operando Pd Faceting-to-Rounding Transitions**

We next used Pd CO oxidation as an experimental benchmark for evaluating whether ChatMOSP can reproduce reaction-temperature-induced nanoparticle restructuring reported by Chee et al.[41] In their operando TEM study, Pd nanoparticles exhibited reversible structural changes during heating (round) and cooling (facet) under $CO/O_2$-containing atmospheres, accompanied by pronounced changes in CO oxidation activity. This system therefore provides a suitable experimental benchmark for testing whether ChatMOSP can reproduce temperature-driven Pd morphology evolution using internally stored MOSP parameters.

Figure 2a illustrates the mobile conversational workflow for the Pd simulation. Through the mobile interface, the user requested Pd nanocluster structures under CO oxidation conditions. ChatMOSP identified the request as an MSR morphology-reconstruction task, retrieved the complete Pd-CO/O parameter set from the built-in database, including facet-dependent surface energies, CO/O adsorption energies, and adsorbate interaction parameters. After displaying the default simulation conditions for user inspection, ChatMOSP allowed the user to modify the temperature and request Pd cluster structures over a series of temperatures. The corresponding MSR input files were then generated automatically, and the reconstructed Pd nanocluster structures were returned directly through the mobile interface. More complete mobile-terminal interactions and output details are shown in Figure S6.

The resulting temperature-dependent Pd morphologies capture the key experimental sequence (Figure 2b). At 473 K, below the experimentally reported transition regime of approximately 400 °C, the Pd nanocluster retains a more faceted morphology with well-defined (111) and (100) low-index surface regions and sharper edges. This low-temperature morphology is consistent with the CO-stabilized faceted state observed by Chee et al.,[41] where adsorbed CO stabilizes low-index Pd facets and suppresses the population of under-coordinated active sites.

When the temperature is increased to 773 K, above the transition regime, the particle loses much of its facet definition and evolves toward a rounded morphology, indicating that the CO-induced facet stabilization is weakened at elevated temperature. At 873 K, the cluster approaches a near-spherical shape with further reduced facet anisotropy, corresponding to the active high-temperature morphology associated with CO oxidation ignition in the operando TEM study. When the temperature is decreased back to 573 K, the faceted morphology partially reforms, consistent with the reversible restructuring observed experimentally during cooling. Therefore, the database-guided ChatMOSP workflow does not merely generate plausible nanoparticle shapes; it reproduces the experimentally observed temperature-induced faceting-to-rounding transition from reaction conditions specified through the mobile interface.

Beyond English text input, ChatMOSP supports multilingual and voice-based interaction. Using the Pd CO oxidation workflow as an example, we demonstrated that the same database-driven MSR simulation can be initiated by Chinese text input and by voice input in both Chinese and English. Representative interaction records, model-comparison results, and corresponding morphology outputs are provided in Figure S6.

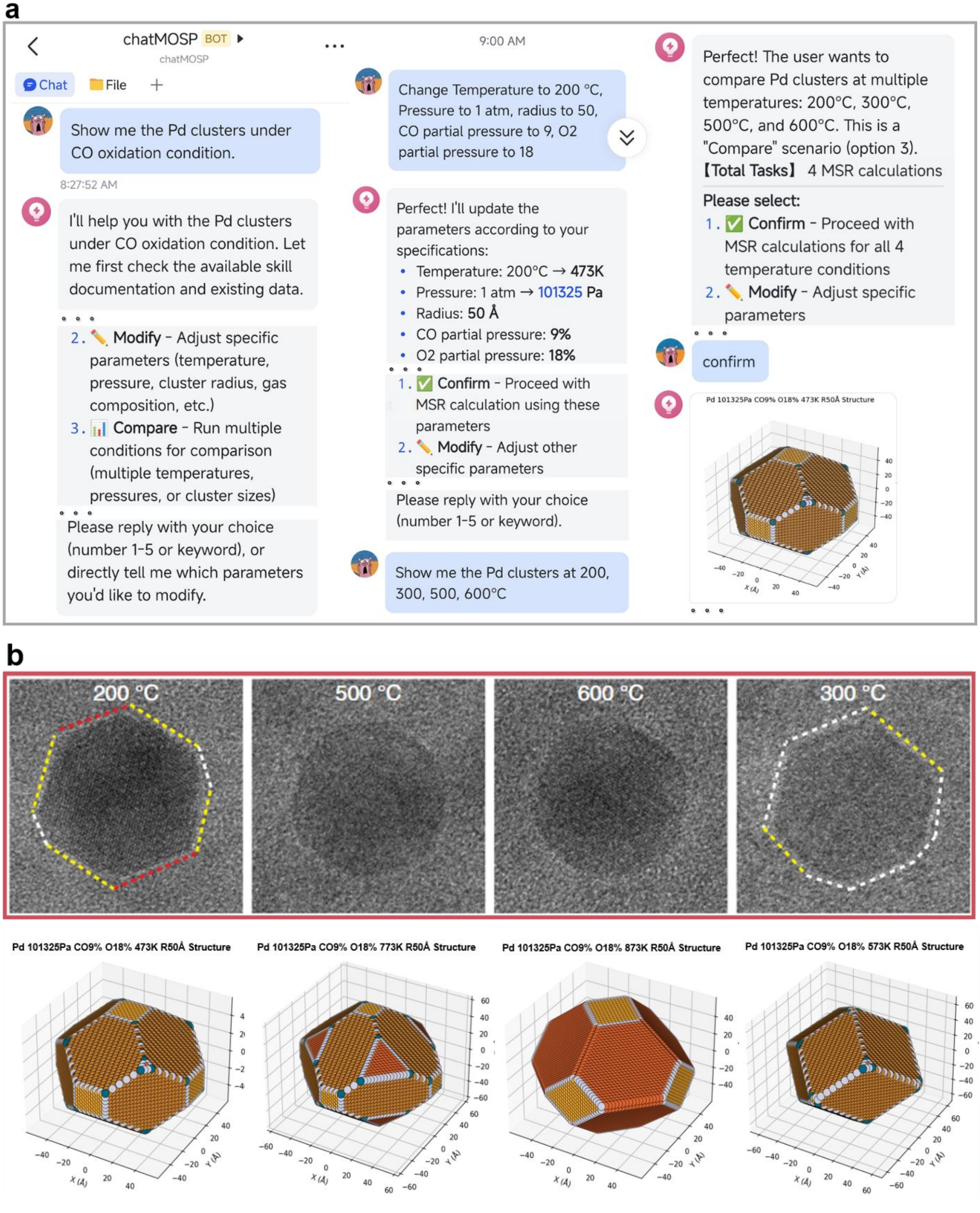

**Figure 2. a.** Mobile interface dialogue showing parameter display and modification for Pd cluster simulation. **b.** Temperature-dependent morphological evolution of Pd clusters from 473 K to 873

K. The morphological trend shows excellent agreement with experimental TEM observations from Chee et al.[41]

**Recovering Pd Morphology Trends through Literature-Assisted Parameterization**

A practical limitation of reaction-condition-dependent morphology simulation is that complete MOSP-ready parameter sets are not always available for a user-selected catalyst and gas environment. To test whether ChatMOSP can move beyond built-in database retrieval, we deliberately removed the Pd-CO/O adsorption energies and adsorbate interaction parameters from the internal MOSP database while keeping the user request unchanged. This created an incomplete-parameter scenario that mimics a common situation in which the experimental reaction conditions are clear, but the corresponding simulation parameters are not directly prepared in MOSP format. Figure 3a exhibits how to use the literature-assisted parameter-completion workflow. The user made the request: "Show me Pd cluster structures under CO oxidation." After detecting the missing entries, ChatMOSP prompted the user to initiate a literature-assisted parameter-construction workflow: "Required parameters are missing from the database. Would you like me to search the literature for these parameters?" Upon user confirmation, the system initiated an automated literature search protocol. In the present implementation, parameter retrieval was performed from online open-access journals. After the user selected the literature-search option and specified a target journal (Nature Communications in this work), ChatMOSP automatically constructed a search query using the metal element combined with the reaction/atmosphere (e.g., "Pd" AND "CO oxidation") and submitted it to the specified journal. The system then listed the top ten results ranked by relevance, performed a preliminary screening of titles and abstracts to filter out alloy studies and unrelated reactions, and recommended several candidate articles. The user could designate a specific article for detailed examination, during which ChatMOSP inspected the main text and downloaded the supplementary information to check for the required parameters; if suitable parameters were identified, they were extracted automatically. The extracted values were then organized into the MOSP parameter format and displayed to the user for inspection before execution. This user-confirmed procedure is important because the literature-retrieval module is intended to assist parameter construction rather than replace expert validation. The detailed search rules, extraction prompts, and parameter-formatting procedures are provided in the Supporting Information.

Using the literature-derived Pd-CO/O parameters, ChatMOSP generated Pd nanocluster morphologies over a temperature series from 548 K to 623 K (Figure 3b). More complete mobile-terminal interactions and output details are shown in Figure S7. At 548 K, the Pd particle exhibits a pronounced faceted morphology, with large flat low-index surfaces (100) and well-defined edges and corners. At 573 K, the Pd particle retains a strongly faceted morphology, with sharp edge/corner features and dominant flat facets. When the temperature is increased to 598 K, the overall faceted shape is preserved, but slight truncation of corners and edges begins to appear. At 623 K, the morphology change becomes more evident: the originally sharp faceted features are weakened, additional truncated surfaces emerge, and the structure evolves toward a more rounded, multi-faceted morphology. Although the absolute transition temperature may depend on the extracted energetic values and model assumptions, the qualitative faceting-to-rounding evolution is consistent with both the database-guided simulation and the operando trend. Thus, even the original parameters were missing, ChatMOSP can automatically retrieve the literature to reproduce the temperature-dependent morphology evolution trend, demonstrating its ability to extend MOSP simulations beyond built-in database coverage by converting literature-derived energetic information into executable, physically constrained simulation inputs.

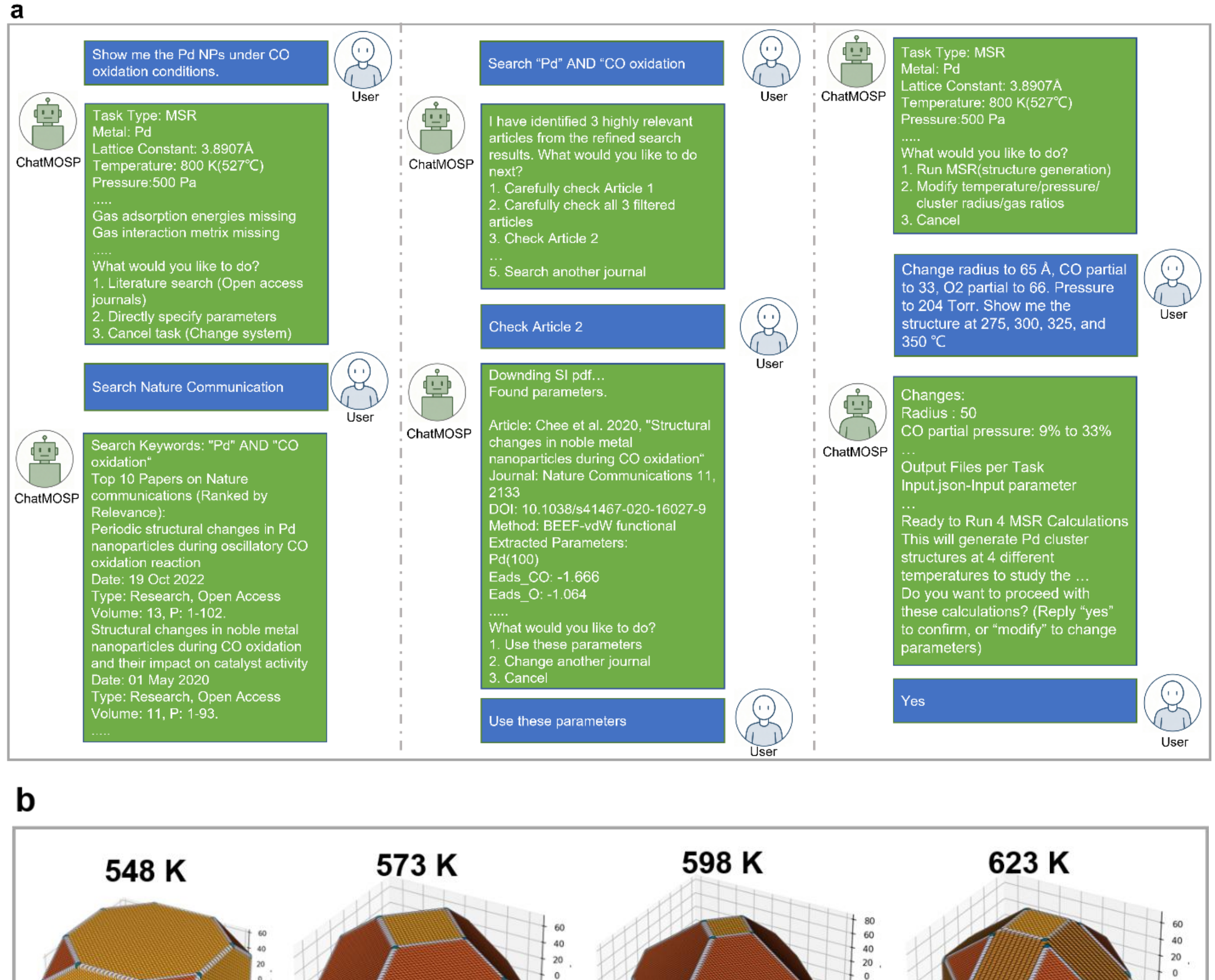

**Figure 3. a.** Mobile interface showing the parameter retrieval workflow. The system identifies missing parameters and presents options for online literature search. **b.** Final Pd cluster structure obtained using literature-derived parameters, showing a morphology trend similar to the experimentally observed faceting-to-rounding evolution.

## End-to-end Full-Scale Study to Revealing Physical Mechanism in Oscillatory Pt CO Oxidation

Oscillatory behavior is a well-known manifestation of nonlinear dynamics in heterogeneous catalysis, arising from feedback among surface reaction kinetics, adsorbate coverage, transport processes, and dynamic catalyst restructuring.[42] In the nanoreactor study by Vendelbo et al., the oscillation in CO conversion was directly visualized together with reversible refacetting of Pt nanoparticles, and the morphology changes may be linked to spatially varying local CO pressures.[13] This observation raises an interesting question for operando catalyst simulation: can a mobile scientific agent translate local reaction conditions into pressure-dependent nanoparticle morphology, catalytic activity, and a physically interpretable feedback mechanism?

We then used ChatMOSP to examine whether the local-pressure-dependent refacetting observed in oscillatory Pt CO oxidation could be captured through a mobile MSR-KMC workflow. As shown in Figure 4a, the user initiated the task by requesting Pt cluster structures under CO oxidation conditions, after which ChatMOSP identified the MSR task, retrieved the required Pt surface, CO/O adsorption, and interaction parameters from the internal MOSP database, and generated reconstructed Pt nanocluster morphologies at 850 K under two CO partial pressures, 150 Pa and 1500 Pa. To further connect morphology with catalytic performance, ChatMOSP subsequently switched to the KMC workflow for the reconstructed structures, displayed the corresponding kinetic parameters, and launched the simulations after user confirmation. After approximately 12 h for 20 million KMC steps, the TOF results were automatically delivered to the mobile device, completing a closed workflow from natural-language input to pressure-dependent Pt morphology and catalytic activity output (Figure 4b).

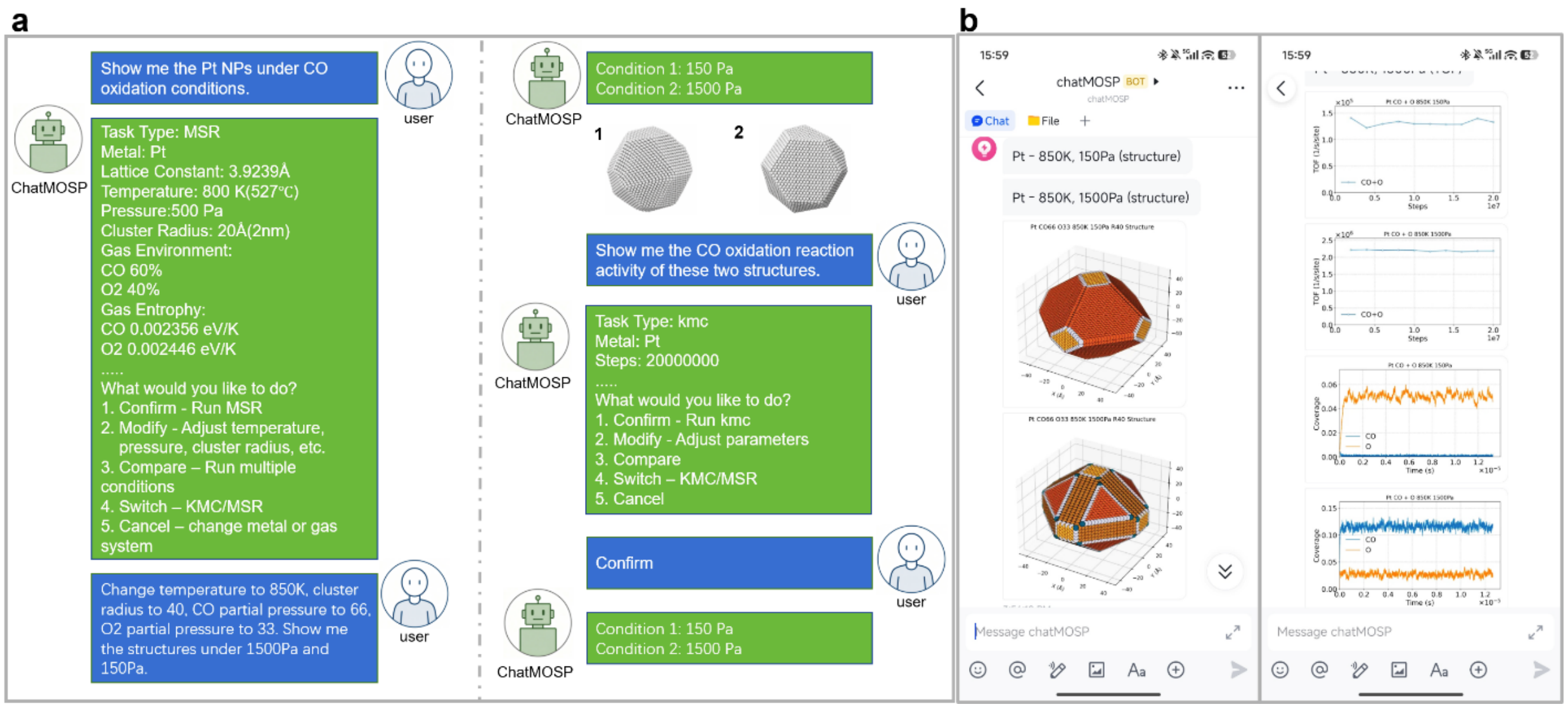

**Figure 4. a.** Mobile conversational workflow for database-driven Pt MSR/KMC simulations **b.** Mobile delivery of ChatMOSP-generated Pt nanocluster morphologies and TOF results.

Figure 5a illustrates the morphology-driven activity-oscillation cycle. At 850 K and 150 Pa CO, ChatMOSP reconstructed a Pt nanocluster with a more faceted morphology, characterized by extended low-index (111) and (100) surface regions, consistent with the experimentally observed morphology under low local CO pressure. The corresponding KMC-derived TOF is markedly lower, indicating a low-activity state. When increasing the CO partial pressure to 1500 Pa, representing a locally CO-rich state, the reconstructed Pt particle became more rounded with reduced facet definition, corresponding to the high-CO-pressure morphology observed experimentally. The corresponding KMC simulation shows that the turnover frequency (TOF) under 1500 Pa CO is more than 10 times higher than that under 150 Pa CO, indicating that the rounded morphology formed under CO-rich conditions represents a high-activity state. Together, these two pressure conditions establish a simulation-based link between local CO pressure, Pt nanoparticle refacetting, and catalytic activity, providing the basis for analyzing the origin of the oscillatory behavior. Additional details of MSR and KMC simulations from mobile-terminal outputs are provided in Figure S8.

After generating the MSR and KMC outputs, we further considered whether ChatMOSP could assist in interpreting the origin of the pressure-induced oscillatory behavior. We therefore entered a series of mobile prompts based on working hypotheses motivated by the experimental observations and the simulated morphology-activity trends. Specifically, we hypothesized that the rounded Pt morphology under high local CO pressure may be associated with higher CO coverage and that the enhanced activity in this state may promote local CO depletion, reduced coverage, and a subsequent transition toward a more faceted particle; this hypothesis was then provided to ChatMOSP as part of the prompt input. On the basis of its own simulation outputs, ChatMOSP was then asked to analyze this possible feedback loop and to render a schematic of the resulting oscillation mechanism. Notably, guided by prompts derived from these working hypotheses, ChatMOSP further reasoned through the possible feedback mechanism and generated both a mechanistic analysis flowchart and a schematic illustration of the oscillatory behavior (Figure 5b). The detailed prompts and returned analyses are provided in Figure S9. In its analysis, high local CO pressure increases CO coverage and favors a rounded Pt morphology enriched with more reactive open sites, resulting in a high-activity state. The enhanced CO oxidation rate then accelerates local CO consumption, lowering the CO pressure and weakening the coverage-driven stabilization of the rounded morphology. As the local CO pressure decreases, the particle shifts back toward a more faceted low-index structure with lower activity, allowing CO to accumulate again and restart the cycle. ChatMOSP attributed this oscillation to three coupled factors: (i) the CO binding energy difference between the (110) surface (-1.865 eV) and the (111) surface (-1.494 eV), with the 25% stronger binding on (110) serving as the primary driver; (ii) surface reconstruction hysteresis, which creates a memory effect that stabilizes the oscillation; and (iii) a reaction-rate feedback loop in which the self-limiting nature of CO consumption drives the system back toward the low-activity state. It further articulated a four-stage oscillation cycle: a low-CO, low-activity state dominated by faceted low-index surfaces at 150 Pa; CO accumulation that raises the local pressure and promotes morphology evolution; a high-CO, high-activity state at 1500 Pa with increased open-site exposure and CO coverage rising from 0.07% to 11.8%; and rapid CO consumption that decreases the local pressure and enables the particle to return toward the faceted morphology. This ChatMOSP-derived mechanism is consistent with the experimental interpretation that oscillatory CO conversion is synchronized with reversible Pt nanoparticle refacetting. Thus, ChatMOSP does more than make MOSP easier to operate: it enables a concrete catalytic phenomenon to be interrogated through a closed chain of condition-dependent morphology prediction, activity simulation, and physically interpretable feedback analysis.

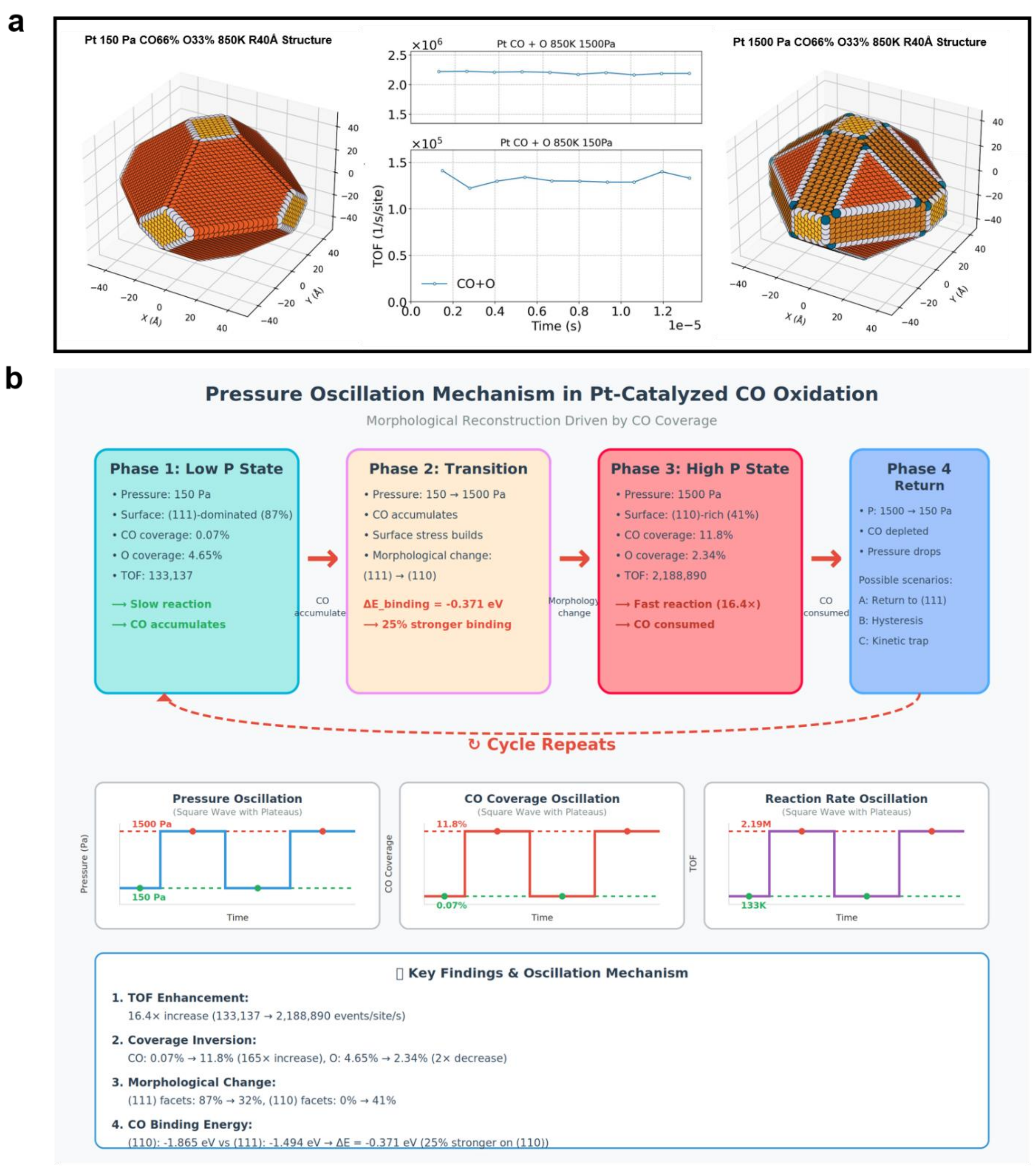

**Figure 5. a.** The ChatMOSP-generated Pt morphologies and KMC-derived TOFs at 850 K under 1500 and 150 Pa CO partial pressures, representing the locally CO-rich rounded high-activity state and the locally CO-depleted faceted low-activity state, respectively. **b.** ChatMOSP-generated

schematic illustration of the morphology-activity-oscillation cycle driven by local CO-pressure switching.

## Conclusions

Overall, ChatMOSP establishes a chemistry-grounded mobile agent workflow for translating experimentally specified reaction environments into physics-based catalyst working-state simulations. By integrating natural-language and voice interaction with an OpenClaw-based hierarchical skill architecture, ChatMOSP converts catalyst identity, temperature, pressure, gas composition, and target observables into parameter-validated MOSP-based MSR and KMC tasks, while ensuring that morphology and activity outputs are generated by the underlying physical models rather than by the language model itself. Using CO oxidation as a representative reaction environment, ChatMOSP captured the temperature-induced facet-to-round evolution of Pd nanoparticles observed by operando TEM, and retained the same morphology trend after Pd-CO/O parameters were removed from the internal database and reconstructed through a literature-assisted workflow. More importantly, for Pt CO oxidation, ChatMOSP linked local CO pressure to adsorbate-stabilized morphology and KMC-derived activity, revealing a pressure-coverage-morphology-activity feedback loop consistent with the experimental interpretation of oscillatory CO conversion. These results show that ChatMOSP does more than lower the operational barrier of MOSP: it transforms a specialized morphology and kinetic simulation package into an accessible, mobile, and physically interpretable tool for analyzing how reaction environments regulate working catalyst structures and catalytic performance.

ASSOCIATED CONTENT

**Supporting Information**.

The supporting information is available free of charge at.

AUTHOR INFORMATION

**Corresponding Author**

**Rui Qi** - Photon Science Research Center for Carbon Dioxide, Shanghai Advanced Research Institute, Chinese Academy of Sciences, Shanghai 201210, China; State Key Laboratory of Low Carbon Catalysis and Carbon Dioxide Utilization, Shanghai Advanced Research Institute, Chinese Academy of Sciences, Shanghai 201210, China; qir@sari.ac.cn

**Beien Zhu** - Photon Science Research Center for Carbon Dioxide, Shanghai Advanced Research Institute, Chinese Academy of Sciences, Shanghai 201210, China; State Key Laboratory of Low Carbon Catalysis and Carbon Dioxide Utilization, Shanghai Advanced Research Institute, Chinese Academy of Sciences, Shanghai 201210, China; Email: zhube@sari.ac.cn

**Yi Gao** - Photon Science Research Center for Carbon Dioxide, Shanghai Advanced Research Institute, Chinese Academy of Sciences, Shanghai 201210, China; State Key Laboratory of Low Carbon Catalysis and Carbon Dioxide Utilization, Shanghai Advanced Research Institute, Chinese Academy of Sciences, Shanghai 201210, China; Email: gaoyi@sari.ac.cn

**Authors**

**Sanyang Ye** - Key Laboratory of Interfacial Physics and Technology, Shanghai Institute of Applied Physics, Chinese Academy of Sciences, Shanghai 201800, China; University of Chinese Academy of Sciences, Beijing 100049, China;

**Author Contributions**

R.Q. initiated the project. B.Z. and Y.G. supervised the project. S.Y. performed the calculations and data analysis. R.Q. wrote the original version. B.Z. and Y.G. revised the manuscript. All authors participated in the discussions.

**Notes**

The authors declare no competing interests.

**ACKNOWLEDGMENT**

This work is supported by National Natural Science Foundation of China (92477105, 92577120), Shanghai Municipal Science and Technology Major Project, and Foundation of the Key Laboratory of Low-Carbon Conversion Science & Engineering, Shanghai Advanced Research Institute, Chinese Academy of Sciences (KLLCCSE-202201Z, SARI, CAS). R. Q. thanks for the Innovation Program of Shanghai Advanced Research Institute, CAS (2025CP007). All calculations were performed at National Supercomputing Center in Shanghai.